# Broadening of plasmonic resonance due to electron collisions with nanoparticle boundary: a quantum-mechanical consideration

Alexander V. Uskov, Igor E. Protsenko, N. Asger Mortensen, Eoin P. O'Reilly

**Abstract.** We present a quantum mechanical approach to calculate broadening of plasmonic resonances in metallic nanostructures due to collisions of electrons with the surface of the structure. The approach is applicable if the characteristic size of the structure is much larger than the de Broglie electron wavelength in the metal. The approach can be used in studies of plasmonic properties of both single nanoparticles and arrays of nanoparticles. Energy conservation is insured by a self-consistent solution of Maxwell's equations and our model for the photon absorption at the metal boundaries. Consequences of the model are illustrated for the case of spheroid nanoparticles and results are in good agreement with earlier theories. In particular, we show that the boundary-collision broadening of the plasmonic resonance in spheroid nanoparticles can depend strongly on the polarization of the impinging light.

A. V. Uskov
P.N. Lebedev Physical Institute, Leninsky pr. 53, 119991, Moscow, Russia;
DTU Fotonik, Technical University of Denmark, Denmark
Plasmonics LTD, Moscow, Russia
email: alexusk@lebedev.ru

I. E. Protsenko
P.N. Lebedev Physical Institute, Moscow, Russia;
Plasmonics LTD, Nijnie Polia Street, 52/1, 119382, Moscow, Russia

N. A. Mortensen
DTU Fotonik, Technical University of Denmark, Ørsteds Plads 343, DK-2800 Kgs. Lyngby, Denmark;
Center for Nanostructured Graphene (CNG), Technical University of Denmark, DK-2800 Kgs. Lyngby, Denmark

E. P. O'Reilly
Tyndall National Institute and Department of Physics, University College Cork, Ireland

**Keywords** Plasmonics nanoparticles • Plasmonic resonance broadening • Surface electron collisions

## Introduction

The electromagnetic properties of sub-wavelength metallic nanoparticles are commonly understood within Claussius-Mossotti theory where the Frölich resonance condition is independent of the particular particle size and shape. Somewhat paradoxically, in this quasi-static limit the resonance frequency and linewidth appear fully determined by the metal bulk properties [1]. On the other hand, it is experimentally well-known (see [2-5] and references therein) that the electromagnetic response reflects the characteristic size $L_{nano}$ of the nanoparticles. As an example, silver nanoparticles exhibit pronounced blueshifts of the resonance frequency in sub-10 nanometer particles [6]. Likewise, the resonance spectral width $\Gamma_{res}$ also depends on the characteristic size $L_{nano}$ of the nanoparticles, indicating the importance of effects beyond the plasmonic bulk properties. Following Kreibig and Vollmer [2], the effects of nanoparticle size on $\Gamma_{res}$ can be divided into two types: (a) *extrinsic* broadening, and (b) *intrinsic* broadening (see Table 2.1 in [2]). *Extrinsic* broadening dominates in relatively large nanoparticles, and is due to electrodynamic effects linked to radiative (dipole) losses in the nanoparticles. *Intrinsic* broadening occurs in relatively small nanoparticles, and is related to the specific behavior of electrons near to the nanoparticle *surface* (boundary). Theories of *intrinsic* broadening are based mainly on two approaches:

(1) using various approximations, the "surface" effects are *included* into an effective ("averaged over

nanoparticle") permittivity $\varepsilon_{eff}$ of the nanoparticle material which is dependent on the nanoparticle size $L_{nano}$: $\varepsilon_{eff} = \varepsilon_{eff}(\omega, L_{nano})$. By construction, the permittivity converges to the bulk permittivity $\varepsilon_{bulk}(\omega)$ of the metal as the particle dimension is increased: $\varepsilon_{eff}(\omega, L_{nano} \to \infty) = \varepsilon_{bulk}(\omega)$. Using $\varepsilon_{eff}$ in Maxwell's equations, we can calculate the electromagnetic response of the nanoparticles, and, in particular, the width $\Gamma_{res}$. In particular, $\varepsilon_{eff} = \varepsilon_{eff}(\omega, L_{nano})$ can be substituted into formulas incorporating Mie's theory for spherical nanoparticles.

(2) the electromagnetic response of nanoparticles is calculated directly from equations-of-motion for an ensemble of electrons interacting with an external electromagnetic field.

These two approaches to describe the *intrinsic* broadening have been developed in the frame of both *classical* and *quantum* descriptions of electrons in nanoparticles. One should note that these approaches both in their classical and quantum versions could in principle be applied to a variety of shapes, although they are in practice generally restricted to more simple nanoparticle shapes such as spheres and cubes (see Table 2.13 in [2]). These theories suffer also from further disadvantages. In particular, pure classical theories are unable to account for photoemission of electrons from nanoparticles beyond a phenomenological level. Quantization of the motion of a huge number of electrons in relatively large nanoparticles can lead to quite cumbersome summing over quantum states even in the case of the relatively simple nanoparticle shapes with a well-known system of discrete energy levels (cube or sphere) – see, for instance, [7].

In the present paper, we proceed along the first line (1), developing an approach to calculate the correction $\Delta\varepsilon''_{surf}(\omega, L_{nano})$ to the imaginary part of the bulk permittivity $\varepsilon_{bulk}(\omega)$ of a metallic nanoparticle due to photon absorption by metal electrons during their collisions with the nanoparticle surface (boundary), and including subsequent electron photoemission from the metal:

$$\varepsilon_{eff}(\omega, L_{nano}) = \varepsilon_{bulk}(\omega) + i\Delta\varepsilon''_{surf}(\omega, L_{nano}) \quad (1)$$

To make further progress, we will assume that the de Broglie electron wavelength $\lambdabar$ in the metal is much shorter than the nanoparticle size $L_{nano}$: $\lambdabar \ll L_{nano}$. More precisely, $\lambdabar$ must be much shorter than the local radius of curvature $R_{curv}$ in any place along the nanoparticle surface:

$$\lambdabar \ll R_{curv} \quad (2)$$

For most metals commonly employed in plasmonics, such as silver or gold, we note that $\lambdabar \approx 0.5\,\text{nm}$. If the condition (2) is fulfilled, one can conveniently consider the boundary between the nanoparticle and surrounding medium as *flat* at any point on the nanoparticle surface. This considerably simplifies the situation and we may calculate the absorption of photons under collision of electrons with the nanoparticle boundary using methods employed in standard quantum-mechanical theory of electron photoemission from metals with a *flat* surface [8-9]. By subsequently equating this *quantum-mechanically* calculated photon absorption rate to the photon absorption rate due to the correction $\Delta\varepsilon''_{surf}$ to bulk permittivity $\varepsilon_{bulk}(\omega)$, calculated in the framework of *classical electrodynamics*, we can finally derive an equation to determine $\Delta\varepsilon''_{surf}$.

There are various physical reasons leading to photon absorption at the boundary, including the potential barrier as well as the discontinuities of the permittivity and of the electron mass at the nanoparticle surface. As we shall see, all these effects can be taken into account in the theory. The theory can be easily applied both to single nanoparticles and to arrays of coupled nanoparticles. While complex-shaped nanoparticles and arrays of interacting nanoparticles may call for numerical methods, we show how analytical progress is indeed possible for ellipsoids and spheroids where explicit analytical formulas can be derived for the permittivity $\Delta\varepsilon''_{surf}$ (see below).

The theory can be formulated to give the permittivity $\Delta\varepsilon''_{surf}$ in the general formula (1), with the bulk permittivity $\varepsilon_{bulk} = \varepsilon'_{bulk} + i\varepsilon''_{bulk}$ taken, for instance, from experiment. For simplicity, we assume that $\varepsilon_{bulk}$ is given by a Drude expression

$$\varepsilon_{Drude}(\omega) = 1 - \frac{\omega_p^2}{\omega^2 + \gamma_{bulk}^2} + \frac{i\gamma_{bulk}\omega_p^2}{\omega(\omega^2 + \gamma_{bulk}^2)} \approx$$
$$\approx 1 - \frac{\omega_p^2}{\omega^2} + \frac{i\omega_p^2}{\omega^3}\gamma_{bulk} \quad (3)$$

where $\omega_p$ is the plasma frequency of the metal, and $\gamma_{bulk}$ is the damping rate associated with bulk interactions. Our strategy is then to calculate the correction $\gamma_{surf}$ to $\gamma_{bulk}$ with subsequent substitution $\gamma_{bulk} \to \gamma_{bulk} + \gamma_{surf}$ into (3) so that

$$\varepsilon_{eff} = \varepsilon_{Drude}(\omega) + \frac{i\omega_p^2}{\omega^3}\gamma_{surf}(\omega, L_{nano}) \quad (4)$$

At this stage, we note by combining Eq. (3) with Mie's theory of a plasmonic resonance, that the spectral width $\Gamma_{res}$ of the plasmonic resonance equals the damping rate $\gamma = \gamma_{bulk} + \gamma_{surf}$ and consequently $\Gamma_{res} = \gamma_{bulk} + \gamma_{surf}$ [2]. In other words, calculating $\gamma_{surf}$, we have also computed the additional broadening of the plasmonic resonance.

In general $\gamma_{surf}(L_{nano})$ must vanish in the bulk limit and it must increase with an increasing Fermi velocity $v_F$. Without referring to a particular microscopic model, it is clear that in a power series of $\gamma_{surf}(L_{nano})$ the leading term must be of the form [2]

$$\gamma_{surf} = A \frac{v_F}{L_{nano}} \qquad (5)$$

where at this stage $A$ is a phenomenological dimensionless constant. Equation (5) includes explicitly the experimentally observed dependence $\gamma_{surf} \propto 1/L_{nano}$ (see [2]). Of course, $A$ can be determined by turning to a particular microscopic model. In general, $A$ must depend on the particular nanoparticle shape as well as material parameters of the metal and the surrounding dielectric matrix. Consequently, $A$ is a basic parameter to calculate in any theory of broadening of plasmonic resonances and it also allows for a direct comparison between different theories [2].

Below we present the method to calculate the parameter $\Delta\varepsilon''_{surf}(\omega, L_{nano})$ for arbitrary nanoparticle shape or for arrays of interacting nanoparticles, and for any general set of material parameters. We illustrate with the method through the example of a spheroid with *infinite* potential barrier at the nanoparticle surface, giving simple analytical formulas for the parameter $A$. The remaining part of the paper is organized as followings. In Sect. 1, the general formulas for the theory are given, and then in Sect. 2 formulas and results for spheroid nanoparticles are presented. Finally, we offer our conclusions.

## 1. Theory

### 1.1. *Electrodynamics formulas*

The following problem is under consideration – see Fig.1. A plane light wave of frequency $\omega$ and with intensity $S$ is propagating along the $z'$-axis in a background matrix with relative permittivity $\varepsilon_e$. The wave is incident on an imbedded metal nanoparticle with relative permittivity $\varepsilon_i = \varepsilon_{eff} = (\varepsilon'_{bulk} + i\varepsilon''_{bulk}) + i\Delta\varepsilon''_{surf}$. The electric field $\mathbf{E}_o$ is polarized along the $y'$ direction. Solving Maxwell equations, for instance, in the quasistatic approximation [1-2,10], one can find the field $\mathbf{E}_i(\mathbf{r})$ inside the nanoparticle. Obviously, $\mathbf{E}_i(\mathbf{r})$ is related to the incident electric field $\mathbf{E}_o$ with a linear relationship

$$\mathbf{E}_i(\mathbf{r}) = \hat{F}(\mathbf{r}) \cdot \mathbf{E}_o \qquad (6)$$

where $\hat{F}(\mathbf{r})$ is an operator (matrix) which depends on the shape and size of the particle, the dielectric constants $\varepsilon_e$ and $\varepsilon_i$, and the light frequency $\omega$. One should stress that the field $\mathbf{E}_i(\mathbf{r})$ [and the operator $\hat{F}(\mathbf{r})$] in (6) depend on the imaginary part $\Delta\varepsilon''_{surf}$ which is as yet unknown, and which we want to find from quantum-mechanical calculations of photon absorption associated with electron collisions at the nanoparticle boundary. We stress that although we are addressing a *single* nanoparticle in the formulation of the problem, the approach developed can in its general form be applied without further complication to an arbitrary nanoplasmonic structure, and in particular, to 2D-arrays of nanoparticles, such as those used in Schottky photodetectors [11-13].

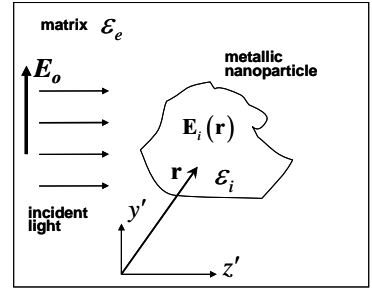

Fig. 1. Schematic illustration of a metallic nanoparticle (with permittivity $\varepsilon_i$) imbedded in a dielectric matrix (with permittivity $\varepsilon_e$). The incident plane wave with electrical field $\mathbf{E}_o$ causes an electric field $\mathbf{E}_i(\mathbf{r})$ inside the nanoparticle.

The power $W$ absorbed in the metal nanoparticle with permittivity $\varepsilon_i = \varepsilon'_{bulk} + i(\varepsilon''_{bulk} + \Delta\varepsilon''_{surf})$ is given as [14]

$$W = 2\omega\varepsilon_o \left(\varepsilon''_{bulk} + \Delta\varepsilon''_{surf}\right) \cdot \int_{volume} d\mathbf{r} \cdot |\mathbf{E}_i(\mathbf{r})|^2 \qquad (7)$$

where the integral is calculated over the nanoparticle volume. Obviously, the contribution from electrons colliding with the nanoparticle boundary is

$$\Delta W = 2\omega\varepsilon_o \cdot \Delta\varepsilon''_{surf} \int_{volume} d\mathbf{r} \cdot |\mathbf{E}_i(\mathbf{r})|^2 \qquad (8)$$

Using (6), we can rewrite (8) as

$$\Delta W = 2\varepsilon_o \cdot \Delta\varepsilon''_{surf} \cdot K_{vol} |\mathbf{E}_o|^2 V_{nano} \qquad (9)$$

where

$$K_{vol} = \frac{1}{|\mathbf{E}_o|^2 \cdot V_{nano}} \cdot \int_{volume} d\mathbf{r} \cdot |\hat{F}(\mathbf{r}) \cdot \mathbf{E}_o|^2 \qquad (10)$$

is a dimensionless coefficient which depends, generally speaking, on the shape and size of the nanoparticle, on the dielectric permittivities $\varepsilon_e$ and $\varepsilon_i$, the light frequency $\omega$, and on the polarization of the incident field $\mathbf{E}_o$ relative to the nanoparticle. Finally, $V_{nano}$ is the nanoparticle volume.

### 1.2. Quantum-mechanical calculation of the photon absorption rate due to electron collisions with the boundary

If the de Broglie electron wavelength $\lambdabar$ in the metal is much smaller than the characteristic nanoparticle size $L_{nano}$, $\lambdabar \ll L_{nano}$, we can safely neglect quantum-confinement effects in the metal. In other words, the electron gas is uniformly distributed with an equilibrium density given by that of the bulk metal. Furthermore, we can calculate the rate $u(\mathbf{r})$ of photon absorption per unit area of nanoparticle surface $[1/(s \times m^2)]$, by considering the nanoparticle surface at the coordinate $\mathbf{r}$ as being *flat*, and by using the theory of photon absorption due to collisions of metal electrons with a *flat* boundary. With this approximation, the rate $u(\mathbf{r})$ is proportional to the square of the normal component $\mathbf{n}(\mathbf{r}) \cdot \mathbf{E}_i(\mathbf{r})$ of the field $\mathbf{E}_i(\mathbf{r})$ [8-9, 15]:

$$u(\mathbf{r}) = C_u \cdot |\mathbf{n}(\mathbf{r}) \cdot \mathbf{E}_i(\mathbf{r})|^2 \qquad (11)$$

The coefficient $C_u$ is calculated quantum-mechanically (see below) and depends, in particular, on the electron density in the metal, on the potential barrier for electrons at the nanoparticle boundary, and on any discontinuities in the permittivity and the electron mass at the interface between the metal and the surrounding medium. Correspondingly, the rate of photon absorption due to electron collisions with the total nanoparticle surface is

$$R_{surf} = \int_{surface} ds\, u(\mathbf{r}) = C_u \int_{surface} ds\, |E_i^{(n)}(\mathbf{r})|^2 \qquad (12)$$

where the integral extends over the entire nanoparticle surface. Using (6), we have

$$R_{surf} = C_u \cdot K_{surf} S_{nano} |\mathbf{E}_o|^2 \qquad (13)$$

where

$$K_{surf} = \frac{1}{S_{nano} |\mathbf{E}_o|^2} \cdot \int_{surface} ds \left| \left( \hat{F}(\mathbf{r}) \cdot \mathbf{E}_o \right)^{(n)} \right|^2 \qquad (14)$$

is the dimensionless coefficient which depends, generally speaking, on the shape and size of the nanoparticle as well as on the dielectric constants $\varepsilon_e$ and $\varepsilon_i = \varepsilon_{bulk} + i\Delta\varepsilon''_{surf}$. Here, $S_{nano}$ is the area of the nanoparticle surface and $\left( \hat{F}(\mathbf{r}) \cdot \mathbf{E}_o \right)^{(n)}$ is the normal component of the vector $\left( \hat{F}(\mathbf{r}) \cdot \mathbf{E}_o \right)$. Naturally, the coefficient is frequency dependent and it is in general sensitive to the polarization of the incident field $\mathbf{E}_o$.

The coefficient $C_u$ in (11) can be found by solving the 1D quantum-mechanical problem for the collision of a single electron with a metal boundary, and then subsequently summing over all metal electrons undergoing such collisions with the surface. To see this in more detail, we first consider an electron plane wave in the metal incident on the metal boundary with wave vector $\mathbf{k}_i$.

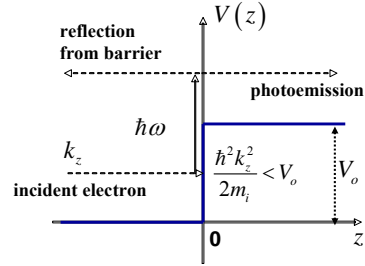

Fig. 2. Schematic illustration of inelastic scattering of an electron at a metal boundary in the presence of an optical field. The potential energy profile $V(z)$ is plotted along the direction $z$ normal to the metal boundary. An electron incident on the boundary (wave vector $k_z$) scatters in-elastically, by absorption of a photon (energy $\hbar\omega$). In the collision with the boundary, the electron can be partly back-reflected into the metal or be forward scattered into the dielectric matrix (photoemission).

In a single-electron description, the work function represents an energy barrier for the electrons at the Fermi level. For simplicity, we will model the equilibrium surface by an abruptly changing potential with a step of height $V_o$, as indicated in Fig. 2. In the absence of an electromagnetic field, the electron scattering is elastic and furthermore the wave vector component parallel to the surface is conserved (since the surface is assumed locally flat). On the other hand, the electron may scatter inelastically in the presence of an electromagnetic field, i.e. by absorbing a photon with energy $\hbar\omega$. While the parallel momentum of the electron is still conserved (neglecting the vanishing momentum of the photon itself), it may either scatter back into the metal or out into the surrounding dielectric matrix. We denote the corresponding scattering probabilities by $p_{in}$ and $p_{out}$, respectively (see Fig.2). Both probabilities $p_{in}$ and $p_{out}$ are proportional to the square of the *normal* to the surface component $\mathbf{n}(\mathbf{r}) \cdot \mathbf{E}_i(\mathbf{r})$ of the field $\mathbf{E}_i$ in the metal [8-9, 15-16]:

$$p_{in} = c_{in} \cdot |\mathbf{n}(\mathbf{r}) \cdot \mathbf{E}_i(\mathbf{r})|^2 \qquad (15)$$

$$p_{out} = c_{out} \cdot |\mathbf{n}(\mathbf{r}) \cdot \mathbf{E}_i(\mathbf{r})|^2 \qquad (16)$$

The complete probability of photon absorption at the surface is

$$p = p_{in} + p_{out} = (c_{in} + c_{out}) \cdot |E_i^{(n)}|^2 \quad (17)$$

The coefficients $c_{in}$ and $c_{out}$ depend on the component $k_{iz}$ of the vector $\mathbf{k}_i = (k_{ix}, k_{iy}, k_{iz})$, the difference $V_o$ between the electron potential energy outside and inside the metal; the shape of the potential $V = V(z)$; the photon energy $\hbar\omega$; the electron masses $m_i$ and $m_e$ in the metal and the surrounding matrix, respectively, and the permittivities $\varepsilon_e$ and $\varepsilon_i$. Obviously, electron photoemission from the metal (i.e., $c_{out} > 0$) can only occur if the electron gains sufficient energy to overcome the work function, i.e. only if $\hbar^2 k_{iz}^2 / (2m_i) + \hbar\omega > V_o$.

The probabilities (15)-(17) can be found using quantum-mechanical methods with appropriate approximations, such as for instance:
− by direct solution of the Schrödinger equation for an electron in the presence of the field using the perturbation theory (see e.g. [15-16]);
− from Fermi's golden rule [8];
− using Green function methods [9].

In passing, we note that for noble metal nanoparticles in vacuum below the plasma frequency, the photon energy is much smaller than the work function and to a good approximation the potential barrier appears effectively infinite (i.e. $V_o = +\infty$). In this very relevant limit the result is particularly simple [15]

$$c_{out} \equiv 0, \quad c_{in} = 4 k_{iz}^2 e^2 / (m_i^2 \omega^4) \quad (18)$$

Summing over all electrons in the metal, colliding with the surface, one can obtain that

$$C_u = \int_{k_{zi} > 0} \frac{2 d\mathbf{k}_i}{(2\pi)^3} \left[ f_F(\mathbf{k}_i) - f_F(\mathbf{k}_f) \right] v_{zi} (c_{in} + c_{out}) \quad (19)$$

where

$$f_F(\mathbf{k}_i) = \left[ 1 + \exp\left( (\varepsilon_i - \varepsilon_F)/k_B T_e \right) \right]^{-1} \quad (20)$$

is the Fermi-Dirac equilibrium distribution function for electrons in the metal, and

$$\varepsilon_i = \hbar^2 (k_{ix}^2 + k_{iy}^2 + k_{iz}^2)/(2m_i)$$

is the (kinetic) energy of the electrons near the Fermi level in the metal. In (19), $v_{iz} = \hbar k_{iz} \cdot m_i^{-1}$ is the normal-component of the velocity for an electron incident on the boundary ($v_{iz} > 0$) and $\mathbf{k}_f = (k_{fx}, k_{fy}, k_{fz})$ is the wave vector of the electron in the metal after photon absorption: $k_{fx} = k_{ix}$, $k_{fy} = k_{iy}$, and $\hbar^2 k_{fz}^2 / (2m_i) = \hbar^2 k_{iz}^2 / (2m_i) + \hbar\omega$. Note that the 3D integrals in (19) can be easily converted into 1D-integrals over $k_{iz}$.

For an infinite barrier $V_o = +\infty$, (18) and (19) combine to give

$$C_u = \frac{2}{3\pi^2} \frac{e^2}{\hbar^2 \omega} \left( \frac{\varepsilon_F}{\hbar\omega} \right)^3 \left[ 1 - \left( 1 - \frac{\hbar\omega}{\varepsilon_F} \right)^3 \right] \quad (21)$$

where the thermal smearing has been assumed much smaller than any other energy scales (justifying a $T_e = 0$ consideration) and furthermore $\hbar\omega < \varepsilon_F$.

### 1.3. Energy balance equation for $\Delta\varepsilon''_{surf}$

Equating the photon absorption power $\Delta W$ due to electron collisions with the nanoparticle boundary (given by (9) in accordance with classical electrodynamics) to the absorption power $\hbar\omega R_{surf}$ (obtained quantum-mechanically – see (13) [1-2]), we now arrive at

$$\Delta\varepsilon''_{surf} = C_u \frac{\hbar}{2\varepsilon_o} \frac{K_{surf}}{K_{vol}} \frac{S_{nano}}{V_{nano}} \quad (22)$$

This is our key result for $\Delta\varepsilon''_{surf}$. One should stress that the coefficients $K_{vol}$ and $K_{surf}$ in (22) are calculated with $\varepsilon_i = \varepsilon_{eff} = \varepsilon_{bulk} + i\Delta\varepsilon''_{surf}$, so that they formally depend on the yet unknown value of $\Delta\varepsilon''_{surf}$. One should remember also that finding $K_{vol}$ and $K_{surf}$ requires solution of Maxwell's equations. The coefficient $C_u$, generally speaking, depends on the difference $(\varepsilon_i - \varepsilon_e)$, and therefore can also depend on $\Delta\varepsilon''_{surf}$. In the next section, we solve (22) for spheroid nanoparticles.

An equation similar to (22) is the central point of many theories, in which the photon absorption rate in nanoparticles is calculated quantum-mechanically with consideration of the metal particle as a potential well (see e.g. [2-5] and references therein). In fact, Eq. (22) is the *energy balance* equation which allows one to solve *self-consistently* Maxwell equations for electromagnetic fields in a plasmonic structure and quantum-mechanical equations of motion for electrons in the metal, including relevant approximations, as appropriate. Such a self-consistent approach guarantees fulfillment of the energy conservation law in the analysis of plasmonics based devices. In particular, with this approach the quantum efficiency of electron photoemission from plasmonic nanoantennas [11-13, 15] is correctly guaranteed not to exceed unity. Proper consideration of such factors can be important especially if resonant effects occur in nanoantenna arrays.

## 2. Broadening of plasmonic resonance in ellipsoid and spheroid nanoparticles

If the nanoparticle has an *ellipsoid* shape with semiaxes $R_a$, $R_b$ and $R_c$, then the field $\mathbf{E}_i(\mathbf{r})$ in the *quasistatic* approximation [1,8] is *homogeneous* inside the nanoparticle, i.e. $\mathbf{E}_i(\mathbf{r}) \equiv \mathbf{E}_i$. Consequently, the operator $\hat{F}(\mathbf{r})$ in (6) does not depend on the coordinate $\mathbf{r}$, i.e. $\hat{F}(\mathbf{r}) \equiv \hat{F}$. On other hand, if the incident wave is polarized along one of the axes of the ellipsoid, the field $\mathbf{E}_i$ is parallel to $\mathbf{E}_o$, so that

$$\mathbf{E}_i = F \cdot \mathbf{E}_o \tag{23}$$

where $F$ is a scalar which can be expressed, in particular, though the polarizability of the nanoparticle [2,10,15], as discussed in the introduction of this paper.

Using (23) in (10) and (14), we obtain for an ellipsoid nanoparticle with light polarized along one of its axes that

$$K_{vol} = |F|^2 \tag{24}$$

$$K_{surf} = |F|^2 \cdot K_g \tag{25}$$

where

$$K_g = \frac{1}{S_{nano}|\mathbf{E}_o|^2} \cdot \int_{surface} ds |(\mathbf{n}(\mathbf{r}) \cdot \mathbf{E}_o)|^2 \tag{26}$$

is a purely geometrical factor, depending only on the nanoparticle shape and the polarization of $\mathbf{E}_o$ relative to the normal vector of the surface.

With (24)-(25), Eq.(22) now becomes

$$\Delta\varepsilon''_{surf} = \frac{\hbar}{2\varepsilon_o} C_u K_g \frac{S_{nano}}{V_{nano}} \tag{27}$$

Quite remarkably, we see how $|F|^2$ drops out and the result carries no trace of the polarization of the particle and the resonance properties themselves. On the other hand, as mentioned above, the quantum-mechanically calculated coefficient $C_u$ can also depend on $\varepsilon_i = \varepsilon_{bulk} + i\Delta\varepsilon''_{surf}$, so that (27) is not yet an equation for $\Delta\varepsilon''_{surf}$. But if we assume that $V_o = +\infty$, then $C_u$ is given by (22), and does not depend on $\varepsilon_i = \varepsilon_{bulk} + i\Delta\varepsilon''_{surf}$ so that in this case (27) then gives the final solution for $\Delta\varepsilon''_{surf}$.

One should stress that this apparent "disappearance of plasmonics" from the balance equation (22) is a consequence of the *quasistatic approximation*. In this limit, plane waves can only excite dipoles in an *ellipsoid* and consequently the field inside the nanoparticle is *homogeneous*. But if we consider a nanoparticle of more complicated shape, or arrays of nanoparticles (even ellipsoids), or work *beyond quasistatics*, when a plane wave can excite also multipole modes even in a sphere, the situation may change. It is obvious that the ratio $K_{surf}/K_{vol}$ in (21) then depends not only on nanoparticle shape, but also on the dielectric constants $\varepsilon_e$ and $\varepsilon_i = \varepsilon_{eff} = (\varepsilon'_{bulk} + i\varepsilon''_{bulk}) + i\Delta\varepsilon''_{surf}$, and can even include a plasmonic resonance or lattice resonance [17] in the case of a 2D array of nanoparticles. These effects in $\Delta\varepsilon''_{surf}$ will be reported elsewhere. Thus, the case considered of an isolated ellipsoid is indeed a special case.

Using the relationship $\Delta\varepsilon''_{surf} = \gamma_{surf}\,\omega_p^2/\omega^3$ in the Drude formula, we can rewrite (27) in the spirit of (5) as

$$\gamma_{surf} = A\frac{v_F}{R_a} \tag{28}$$

with

$$A = F_{mat} \cdot F_{shape} \cdot K_g \tag{29}$$

where the coefficient

$$F_{mat} = \frac{4}{3\pi}\alpha \cdot \frac{c}{v_F} \frac{\omega_p}{\omega} \left(\frac{\varepsilon_F}{\hbar\omega_p}\right)^3 \left[1 - \left(1 - \frac{\hbar\omega}{\varepsilon_F}\right)^3\right] \tag{30}$$

depends only on material parameters of the nanoparticle with $\alpha = e^2/(4\pi\varepsilon_o\hbar c) = 0.007297 \sim 1/137$ being the fine-structure constant. Likewise, the parameter

$$F_{shape} = \frac{R_a S_{nano}}{V_{nano}} \tag{31}$$

depends only on nanoparticle shape. The semiaxis $R_a$ is chosen in (28) as the nanoparticle characteristic size.

If the nanoparticle is a spheroid with semiaxes $R_a = R_b$ and $R_c$ (i.e. the semiaxis of the spheroid along its rotation axis is equal to $R_c$), and with $r_{asp} = R_a/R_c$ as the aspect ratio, then $F_{shape} = F_{shape}(r_{asp})$,

$$F_{shape}(r_{asp}) = 3r_{asp}^{-1}G_c(r_{asp}) \tag{32}$$

with

$$G_c(r_{asp}) = 0.5 \cdot \left[r_{asp}^2 + \left(\sqrt{1 - 1/r_{asp}^2}\right)\tanh^{-1}\left(\sqrt{1 - 1/r_{asp}^2}\right)\right] \tag{33}$$

If the light polarization $\mathbf{E}_o$ is *normal* to the rotation axis of the spheroid,

$$K_g = K_g^\perp(r_{asp}) = 0.5 \cdot H_\perp(r_{asp})/G_c(r_{asp}) \tag{34}$$

with

$$H_\perp = \frac{r_{asp}}{2}\left[\frac{r_{asp}}{1 - r_{asp}^2} + \frac{1 - 2r_{asp}^2}{(1 - r_{asp}^2)^{3/2}} \cdot \sin^{-1}\left(\sqrt{1 - r_{asp}^2}\right)\right] \tag{35}$$

For polarization, *parallel* to the spheroid rotation axis,

$$K_g = K_g^\parallel(r_{asp}) = 0.5 \cdot H_\parallel(r_{asp})/G_c(r_{asp}) \tag{36}$$

$$H_\parallel = r_{asp}^3\left[\frac{-r_{asp}}{1 - r_{asp}^2} + \frac{1}{(1 - r_{asp}^2)^{3/2}} \cdot \sin^{-1}\left(\sqrt{1 - r_{asp}^2}\right)\right] \tag{37}$$

Fig.3 demonstrates $F_{shape}(r_{asp})$, $K_g^\perp(r_{asp})$ and $K_g^\parallel(r_{asp})$. One sees that $K_g^\perp(r_{asp})$ and $K_g^\parallel(r_{asp})$ show very different, but quite expectable behavior with increasing $r_{asp}$: as $K_g^\perp(r_{asp})$ decreases and tends to zero, then $K_g^{\parallel}(r_{asp})$ increases and tends to one. Note that the spheroid becomes a thin disk as $r_{asp} \to \infty$. This substantially affects the magnitude of the coefficient *A* for the two different polarizations – see below. We note also that $K_g^{\perp} = K_g^{\parallel} = 1/3$ and $F_{shape} = 3$ for a sphere ($r_{asp} = 1$), so that for a sphere $A \equiv F_{mat}$.

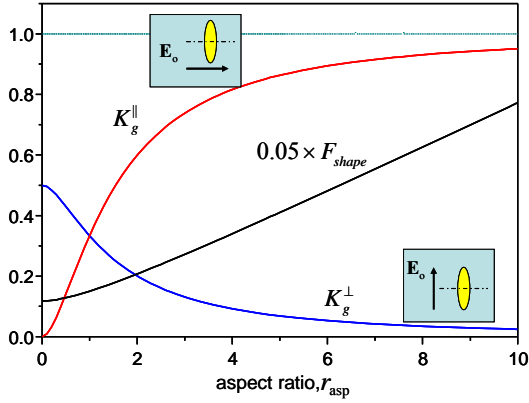

Fig.3. Dependence of $F_{shape}(r_{asp})$, $K_g^{\perp}(r_{asp})$ and $K_g^{\parallel}(r_{asp})$ on the aspect ratio $r_{asp}$.

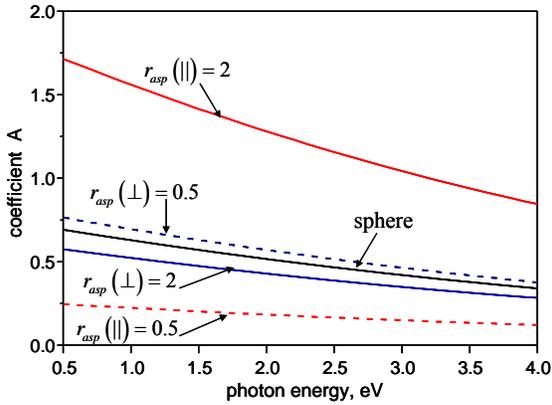

Fig.4. Spectral dependence of *A* for different spheroid shapes and light polarizations relatively to the rotation axis of the spheroid. *Parallel* polarization: red solid is for $r_{asp} = 2$, and red dashed is for $r_{asp} = 0.5$; *normal* polarization: blue solid is for $r_{asp} = 2$ and blue dashed is for $r_{asp} = 0.5$. Black solid – sphere.

Figure 4 demonstrates the spectral dependence of $A(\hbar\omega)$ for various spheroid shapes, namely $r_{asp} = 0.5$, 1 (sphere), and 2. In these calculations, we used the following parameters (appropriate for gold): $\varepsilon_F = 5.51\,\mathrm{eV}$ and $\hbar\omega_p = 8.95\,\mathrm{eV}$. One sees that for the range of photon energies considered (0.5 to 4 eV) the coefficient *A* changes approximately by a factor of 2. In particular, for a sphere *A* changes from 0.69 to 0.34, and for $\hbar\omega = 1\,\mathrm{eV}$ the coefficient is $A = 0.63$. For comparison, the coefficient *A* is equal to 0.75 in the classical model, based on Matthiessen's rule and on diffuse isotropic scattering at the boundary [2].

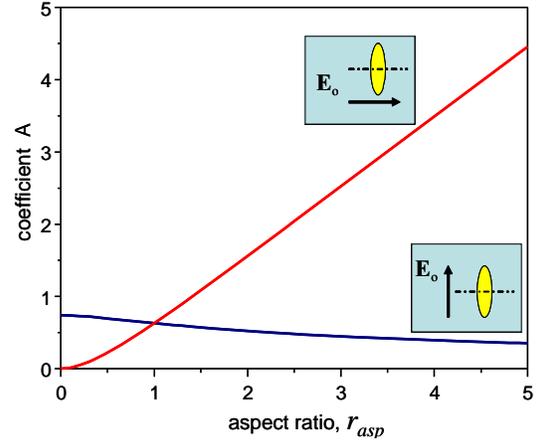

Fig.5. Dependence of *A* on the aspect ratio $r_{asp}$ for different light polarizations for $\hbar\omega = 1\,\mathrm{eV}$: red – polarization is parallel to the rotation axis; blue – polarization is normal to the rotation axis.

Figure 5 shows the dependence of *A* on the aspect ratio $r_{asp}$ for $\hbar\omega = 1\,\mathrm{eV}$. One sees that in contrast to the case of polarization normal to the rotation axis (blue curve) where the coefficient changes not so much, in the case of parallel polarization (red) the coefficient A changes strongly, tending to zero at $r_{asp} \to 0$ and increasing linearly for $r_{asp} \geq 1$. The latter behaviour is explained as follows. If the semiaxis $R_a$ is constant, and the aspect ratio is increasing, so that the semiaxis $R_c$ along the rotation axis is decreasing, the spheroid becomes a thin disk with the field $\mathbf{E}_o$ being normal to the upper and lower faces of the disk. In this case, electrons can effectively collide with the *entire* nanoparticle surface, which is $\sim 4\pi R_a^2$, and therefore the absorbed power does not increase further when $r_{asp} \geq 1$. On the other hand, in the calculation of $\gamma_{surf}$ we ascribe this constant absorption to the nanoparticle volume $V_{nano}$ which is decreasing with increasing $r_{asp}$. Consequently, this requires to have *A* increasing linearly with $r_{asp}$ – see Fig.5. From Fig.5 one sees that for *parallel* polarization the coefficient *A* changes from 0 to ~4.5 if the aspect ratio $r_{asp}$ changes from 0 to 5, but for *normal* polarization *A* changes from 0.77 to 0.34. Thus, broadening of the plasmonic resonance due to electron collisions with the nanoparticle boundary can strongly depend on the light polarization direction if the nanoparticle is not spherical. This conclusion is in agreement with [2,4].

**Conclusion**

We have suggested an approach to calculate the correction to the imaginary part of the metal susceptibility of plasmonic nanostructures due to electron collisions with the nanostructure boundary, and the corresponding broadening of the plasmonic resonances due to such collisions. The approach is suited both to treat a single nanoparticle of arbitrary shape and also to treat an array of nanoparticles. The method works if the curvature of the nanoparticle surface varies slowly on the scale of the de Broglie electron wavelength in the metal. In this case, the quantum-mechanical theory of photon absorption at a flat metal boundary is valid. It is important to stress that the approach allows calculations of the plasmonic resonance broadening without solving the Schrödinger equation even for nanostructures with complicated shapes. One should also note that corrections of the order of $1/R_{curv}$ due to a finite radius of curvature $R_{curv}$ can be included in the theory of photon absorption at metal boundaries [9], and correspondingly can also be included in the method developed here to calculate the plasmonic resonance broadening.

Also while most quantum-mechanical theories concentrate (explicitly or implicitly) on studying the plasmonic resonance broadening of *dipole* modes with an *homogeneous* electric field inside the nanoparticles, our approach can treat the more general case. It allows one to calculate for instance broadening due to multipole resonances, say, in spheres, and also the broadening of plasmonic resonances in 2D lattices of nanoantennas for Schottky photodiodes where there is an *inhomogeneous* field distribution inside the nanostructures. Obviously, this latter example would need substantial numerical efforts.

Because the approach presented here includes a self-consistent solution of Maxwell's equations for the field in the nanostructure and a quantum-mechanical calculation of the photon absorption at nanostructure boundary, it guarantees energy conservation in the calculations. This ensures for example to obtain a quantum efficiency not higher than one for Schottky photodiodes with 2D lattices of nanoantennas with resonance responses of different kinds .

Application of the approach to spheroid and sphere nanoparticles gives analytical expressions whose values are close to the results obtained in various earlier theories.

Finally, we note in the analysis presented here we have for simplicity considered a local-response approximation. Nonlocal response is known to smear the induced surface charges over a finite length scale (of the order of the Fermi wavelength) with according modifications of the electrical field near the surface [18,19]. Within the hydrodynamic model the boundary conditions for the fields are however well-known [18,20] and the formalism developed here can easily be applied to also treat such effects.

**Acknowledgements**

This work was partly supported by Science Foundation Ireland (06/IN.1/I90). The Center for Nanostructured Graphene (CNG) is sponsored by the Danish National Research Foundation, Project DNRF58